\documentclass[preprint,12pt,authoryear]{elsarticle}

\usepackage{amssymb}
\usepackage{amsmath}
\usepackage{amsthm}
\usepackage{multirow}
\usepackage{xcolor}
\usepackage[most]{tcolorbox}
\usepackage{todonotes}
\usepackage{placeins}
\usepackage{booktabs}
\usepackage{tabularx}
\usepackage{rotating}
\usepackage{subcaption}
\usepackage{enumitem}
\usepackage{hyperref}
\usepackage{listings}
\usepackage{xcolor}
\usepackage{longtable}
\newtcolorbox{graybox}{
  colback=gray!20,
  colframe=gray!20,
  boxrule=0pt,
  arc=0pt,
  left=8pt,
  right=8pt,
  top=6pt,
  bottom=6pt,
  boxsep=0pt,
  breakable
}

\usepackage{listings}
\newcommand{\wti}{\textsc{wti}}

\newcommand{\gpt}{GPT-4o }
\newcommand{\llama}{Llama 3.2-3b }
\newcommand{\finbert}{FinBERT }

\begin{document}

















\begin{frontmatter}

\title{Beyond Polarity: Multi-Dimensional LLM Sentiment Signals for WTI Crude Oil Futures Return Prediction}

\author[author1]{Dehao Dai}
\author[author2]{Ding Ma}
\author[author3]{Dou Liu}
\author[author4]{Kerui Geng}
\author[author5]{Yiqing Wang\corref{cor1}}

\cortext[cor1]{Corresponding Author: Yiqing Wang}
\ead{woshilucy712@gmail.com}

\affiliation[author1]{organization={University of California San Diego},
            city={La Jolla},
            state={California},
            country={United States}}
\affiliation[author2]{organization={Georgia Institute of Technology},
            city={Atlanta},
            state={Georgia},
            country={United States}}
\affiliation[author3]{organization={New York University},
            city={New York},
            state={New York},
            country={United States}}
\affiliation[author4]{organization={Tulane University},
            city={New Orleans},
            state={Louisiana},
            country={United States}}
\affiliation[author5]{organization={Southern Methodist University},
            city={Dallas},
            state={Texas},
            country={United States}}

\begin{abstract}
Forecasting crude oil prices remains challenging because market-relevant information is embedded in large volumes of unstructured news and is not fully captured by traditional polarity-based sentiment measures. This paper examines whether multi-dimensional sentiment signals extracted by large language models improve the prediction of weekly \wti{} crude oil futures returns. Using energy-sector news articles from 2020 to 2025, we construct five sentiment dimensions covering relevance, polarity, intensity, uncertainty, and forwardness based on \gpt, \llama, and two benchmark models, \finbert and AlphaVantage. We aggregate article-level signals to the weekly level and evaluate their predictive performance in a classification framework. The best results are achieved by combining \gpt and \finbert, suggesting that LLM-based and conventional financial sentiment models provide complementary predictive information. SHAP analysis further shows that intensity- and uncertainty-related features are among the most important predictors, indicating that the predictive value of news sentiment extends beyond simple polarity. Overall, the results suggest that multi-dimensional LLM-based sentiment measures can improve commodity return forecasting and support energy-market risk monitoring.
\end{abstract}


\end{frontmatter}




\section{Introduction}
\label{SecIntro}

Crude oil remains a cornerstone strategic resource for modern economies and plays a fundamental role in sustaining economic growth and social stability \citep{hamilton1983oil,barsky2002oil, ma_jumps_2020}. As the most actively traded energy commodity, crude oil prices influence inflation, industrial production, transportation costs, and financial market dynamics. 
Accordingly, crude oil futures markets have become the primary venues for price discovery, and accurate forecasting of crude oil price remains a persistent challenge for market participants. There is a considerable literature to address this challenge including traditional statistical frameworks used in high-dimensional time series like ARIMA and GARCH
\citep{xiang2013application,agnolucci2009volatility}, feature selection \citep{zhang_forecasting_2022,song2023forecasting} and modern machine learning techniques such as LSTM networks and deep learning models \citep{zhao2017deep,guo2023forecasting}. These models emphasize quantitative analysis of structured data to provide well-performed crude oil price forecasting and have also been widely applied to broader equity markets. However, their forecasting accuracy is constrained due to lack of the crude oil markets' intrinsic features. 


Different from equity markets, which are largely driven by firm-level fundamentals, crude oil futures markets are highly sensitive to expectation-driven risk factors, including supply disruptions, geopolitical conflicts, policy interventions, and global demand uncertainty \citep{ames_which_2020}. 
In contrast to numerical data, text data like news, reports and social media can provide a new framework to forecast crude oil prices by extracting market signals. These resources encode timely sentiment, capture shifts in expectations, and often contain policy or geopolitical cues that influence oil price dynamics. Recent work on effective crude oil price forecasting using text-based and big-data-driven models further supports this direction \citep{wu2021effective}.


There is a substantial literature evaluating the role of media tone and news sentiment in asset return prediction. Early studies on textual sentiment rely on dictionary-based approaches that quantify the frequency of positive and negative words to measure tone \citep{tetlock_giving_2007,chen2018ntusd}. More recent work employs transformer-based models such as \finbert to classify financial text polarity \citep{yahia2024impact}and forecast both returns and volatility \citep{li2024novel}. Although these tools have improved sentiment measurement accuracy, they primarily focus on directional tone (positive versus negative) and sentiment intensity. However, oil futures pricing is driven not only by directional sentiment but also by forward-looking uncertainty and expectation revision. A news article may be neutral in tone yet signal elevated uncertainty or significant forward guidance regarding future supply conditions. Polarity-based measures are therefore structurally limited in capturing these economically relevant dimensions.

Recent advances in large language models (LLMs) provide an alternative framework for semantic decomposition. Unlike traditional tools, LLMs can extract multi-dimensional attributes from text, including relevance, polarity, intensity, uncertainty, and forward-looking orientation. These dimensions may better align with the theoretical drivers of commodity pricing, \citep{kim2023llms,wang2025novel,jeong2025energy} particularly in markets where geopolitical risk and expectation formation play central roles. Building on this insight, we examine whether LLM-derived signals offer incremental predictive value for \wti{} crude oil futures.

In this papeinvestigates whether multi-dimensional sentiment signals extracted by LLMs provide incremental predictive power for weekly returns of \wti{} crude oil futures. Using energy-related news articles from 2020 to 2025 from AlphaVantage, we construct weekly aggregated sentiment features generated by three models: \gpt, \llama, and \finbert. We compare their predictive performance using a LightGBM classifier and evaluate feature contributions through SHAP-based explainability analysis.

Our findings suggest that LLM-derived uncertainty and forward-looking dimensions exhibit statistically and economically significant predictive power beyond traditional polarity measures. Multi-dimensional
LLM-based sentiment measures can improve commodity return forecasting
and support energy-market risk monitoring.Our findings contribute to the emerging area of LLM applications in financial economics and provide new evidence on the importance of multi-dimensional textual signals in commodity futures forecasting.

The remainder of the paper is organized as follows. Section~\ref{SecLiterature} reviews related work. Section~\ref{SecMethodlogy} describes the end-to-end workflow pipeline from data collection to prediction modeling evaluation. Section~\ref{SecResults} reports empirical results, and Section~\ref{SecConclusion} discusses the results and future directions.

\section{Related Literature}
\label{SecLiterature}

\subsection{Crude Oil Futures Forecasting}

Predicting crude oil price movements has drawn scholars' interest due to the oil market’s economic importance and the naturally high volatility of crude oil prices. Early economic approaches rely on traditional econometric and statistical models, including ARIMA, volatility models, and market-based predictors. Short-term forecasting with ARIMA and linear time series models remains a common benchmark \citep{xiang2013application, nasir2023new}. Earlier volatility-oriented studies, such as \citet{cunado_oil_2005}, highlighted the strong macro-financial linkages surrounding oil price dynamics, and \citet{agnolucci2009volatility} showed that conditional heteroskedastic models provide important insights into the time-varying uncertainty of crude oil futures markets. 

There is also considerable literature that emphasizes machine learning and deep learning methods to capture the strong nonlinearity, regime dependence, and complex maturity structure of oil futures markets. For example, \citet{barunik2016forecasting} modeled the crude oil futures curve through a dynamic Nelson–Siegel framework enhanced by neural networks, showing that nonlinear methods can improve forecasts of the term structure. Focusing on the Chinese market, \citet{guo2023forecasting} and \citet{zhai2025research} further developed the framework of deep learning to exploit historical prices, volatility, and other nonlinear features to enhance crude oil futures forecasting accuracy, while \citet{zhao2017deep} and \citet{guan2023new} proposed deep neural and hybrid learning models in extracting latent patterns from highly volatile oil price series.

Moreover, recent studies adopt hybrid and multi-step forecasting frameworks to improve predictive performance over longer horizons. For instance, \citet{duan2022novel} developed a novel crude oil futures forecasting approach that combines multiple modeling stages, and \citet{chen2018multi} showed that hybrid designs can achieve superior performance in multi-step-ahead prediction by integrating complementary information from different model classes.

A consistent finding across this literature is that energy markets are highly sensitive to expectation-driven shocks, like geopolitical events, OPEC+ policy announcements, and demand uncertainty, which has motivated researchers to integrate news-based information into forecasting models. For example, \citet{cepni2022news} adopted news media and attention spillovers across energy markets to possess significant predictive content for crude oil futures prices; \citet{li2019text} proposed a text-based deep learning approach that exploits information embedded in online news media to improve crude oil price forecasting performance; \citet{sadik2020forecasting} incorporated macroeconomic news into a predictive model for forecasting crude oil futures prices.

\subsection{Sentiment Analysis in Finance}
Sentiment analysis \citep{medhat2014sentiment} has become an important research area in finance to analyze how people's opinions and emotions affect financial decisions. Earlier studies used a traditional ``word count'' approach \citep{guo2016textual, loughran2016textual} to employ sentiment analysis techniques. The most popular word lists used in finance research are the Harvard IV psychosocial word lists \citep{kearney2014textual}, which have also been developed in academic finance \citep{tetlock_giving_2007,tetlock2008more,twedt2012reading}. \citet{bollen_twitter_2011} and \citet{siganos2017divergence} used general dictionaries to test for relationships between sentiment from social media and stock index returns.

In recent years, machine learning has challenged traditional approaches due to the lack of accuracy in dictionary methods \citep{guo2016textual,renault2017intraday,mcgurk2020stock}. A commonly used ML algorithm is the probabilistic Naive Bayesian classifier to detect sentiment in finance \citep{antweiler2004all,li2010information,sprenger2014tweets}. Some other works adopted the Reuters NewsScope Sentiment Engine \citep{gross2011machines,sun2016stock,audrino2019sentiment}. Recent work introduced the transformer architecture \citep{sun2019fine,hiew2019bert,huang2023finbert,lopez2023can,touvron2023llama} to analyze the sentiment in the finance domain, which is also a highlight technique in natural language processing. More discussions can be followed in \citet{todd2024text}.

\subsection{Multi-Dimensional Sentiment and Uncertainty}

The multi-dimensionality of sentiment has been further emphasized by studies distinguishing between rational uncertainty and behavioral sentiment. 
\citet{saravanaraj2025sentimental,kang2025multidimensional,eckhaus2026data} proposed a multidimensional sentiment model to examine the varying influence of investor sentiment on stock market volatility and stock returns. Modern machine learning techniques are also used in multi-dimensional sentiment analysis. For example, \citet{wang2025deep} proposed a deep learning-based model to fuse various investor sentiment information represented through financial texts. 

Uncertainty is also a crucial factor in financial asset pricing. \citet{merton1973intertemporal} introduced the uncertain shifts in investment opportunities into the traditional capital asset pricing model. Under this framework, there are considerable works to study the relationship between uncertainty and pricing \citep{rigotti2005uncertainty,jiang2005information,zhang2006information,arnold2008fundamental,dash2021economic,asgharian2023effect}. Some recent studies have extended previous research. For example, \cite{birru2022sentiment} first investigated the interaction between investor sentiment and uncertainty and introduced the volatility index to measure the dynamic uncertainty in the market. \cite{seok2024dual} concluded that the effect of daily sentiment on short-term returns is more significant when uncertainty is increased. More recent discussions on uncertainty can be found in \citet{arnold2010treasury,segal2015good,manela2017news}.

\section{Methodology}
\label{SecMethodlogy}

\subsection{Data Source and Cleaning}
\label{SubsecData}

We extract energy-sector news articles from the AlphaVantage News Sentiment
API \citep{alphavantage_api}. AlphaVantage is a financial data provider offering market data APIs covering equities, commodities, news, and other information. The available news dataset is accompanied by topic and ticker tags to facilitate filtering. To restrict to crude oil-related content, we only extract news with the \texttt{energy\_transportation} topic tag.

The full sample comprises news articles spanning six years, from 1 January 2020 to 31 December 2025, yielding  29,153  articles after deduplication. To accommodate the API's constraint of 1,000 articles per call, articles were retrieved on a monthly basis and stored in raw JSON format prior to processing. Subject to budget and  computational constraints, we draw a stratified random sample of $20\%$ of the full corpus (8,639 articles) for sentiment analysis extraction. Stratification is performed by calendar month, ensuring uniform temporal coverage across the whole sample window from 2020 to 2025. The same sample is used for all three extraction models to ensure comparability. Table~\ref{tab:data_summary} provides descriptive statistics for the
article corpus.

\FloatBarrier
\begin{table}[htbp]
\centering
\caption{News Article Corpus: Summary Statistics}
\label{tab:data_summary}
\begin{tabularx}{\textwidth}{Xrr}
\hline
 & Full Sample & Sentiment Sample (20\%) \\
\hline
Total number of articles          & 29{,}153  & 8{,}639 \\
Date range              & \multicolumn{2}{c}{01/01/2020--12/31/2025} \\
Avg.\ articles/week     & 93   & 28  \\
\hline
\end{tabularx}
\end{table}

Daily closing prices for the front-month \wti{} crude oil futures contract (\texttt{CL=F}) are obtained from Yahoo Finance via the \texttt{yfinance} Python library.
We compute weekly log returns as:
\begin{equation}
    r_t = \ln\!\left(\frac{P_t}{P_{t-1}}\right),
    \label{eq:weekly_return}
\end{equation}
where $P_t$ denotes the closing price on the last trading day of week $t$.

The full sample spans 1,509 trading days covering 314 weeks. The mean of weekly log return is $0.18\%$ with standard deviation $6.37\%$. The minimum return happened at week 16 March 2020 to 20 March 2020 with $-29.31\%$ log return while the maximum happened at week 30 March 2020 to 30 April 2020 with $31.75\%$ log return. Out of 314 weeks, the first week (12/30/2019 - 01/03/2020) does not yield a valid weekly return, 166 weeks (52.87\%) are classified as up-weeks with positive weekly log return and 147 weeks (46.82\%) as down-weeks with negative values. 

The binary prediction target is defined as:
\begin{equation}
    y_t = \mathbf{1}\!\left[r_{t+1} > 0\right],
\end{equation}
i.e., whether the \wti{} weekly log return in week $t+1$ is positive.
All sentiment features used to predict $y_t$ are constructed from news
articles published no later than the end of week $t$ to prevent
look-ahead bias.




\subsection{Sentiment Extraction}
\label{SubsecSentiment}
We consider three models to extract sentiment signals, including \gpt 
\citep{achiam2023gpt}, \llama \citep{grattafiori2024llama}, and 
\finbert \citep{huang2023finbert}. The three models represent a 
spectrum of architectural approaches. \gpt serves as a 
state-of-the-art instruction-tuned large language model with strong semantic reasoning capabilities; \llama provides a lightweight open-source alternative that enables cost-efficient inference; and \finbert offers a domain-specific encoder fine-tuned on financial text, serving as a competitive supervised baseline.

Given a structured prompt (see  ~\ref{app:prompts}),\gpt and \llama produce five sentiment dimensions for each article: 
\begin{itemize}
    \item \textbf{Relevance}: measures whether an article contains information pertinent to \wti{} crude oil markets or the broader energy sector, ranging from 0 (no relevance) to 1 (highly relevant).
    
    \item \textbf{Polarity}: captures the directional sentiment of an article, reflecting whether the content conveys a bullish or bearish outlook for crude oil prices, ranging from $-1$ (extremely bearish) to $1$ (extremely bullish).  
    
    \item \textbf{Intensity}: quantifies the strength or conviction of the expressed sentiment, distinguishing between mildly and strongly worded articles of the same polarity, ranging from 0 (weak) to 1 (strong).
    
    \item \textbf{Uncertainty}: reflects the degree to which an article expresses ambiguity, risk, or lack of clarity about future market outcomes. It captures hedging language and contested forecasts, ranging from 0 (certain) to 1 (highly uncertain).
    
    \item \textbf{Forwardness}: measures the extent to which an article contains forward-looking language, such as projections, expectations, or policy outlooks, ranging from 0 (past events) to 1 (future outlook).
    
\end{itemize}

\finbert, by contrast, is a classification model fine-tuned for 
sentiment analysis in the financial domain and provides only polarity and intensity scores. The AlphaVantage API supplies a polarity score for each article without additional dimensions. 
Table~\ref{tab:dimensions} provides an overview of the sentiment 
dimensions available from each model.

\FloatBarrier
\begin{table}[htbp]
\centering
\caption{Sentiment Dimensions by Model}
\label{tab:dimensions}
\begin{tabular}{lccccc}
\hline
Model & Relevance & Polarity & Intensity & Uncertainty & Forwardness \\
Range & {[}0, 1{]} & {[}-1, 1{]} & {[}0, 1{]} & {[}0, 1{]} & {[}0, 1{]} \\
\hline
\gpt        & $\checkmark$ & $\checkmark$ & $\checkmark$ & $\checkmark$ & $\checkmark$ \\
\llama   & $\checkmark$ & $\checkmark$ & $\checkmark$ & $\checkmark$ & $\checkmark$ \\
\finbert         &              & $\checkmark$ & $\checkmark$ &              &              \\
AlphaVantage    &              & $\checkmark$ &              &              &              \\
\hline
\end{tabular}
\end{table}

Formally, for article $i$ published in week $t$, each model produces a
score vector:
\begin{equation}
    \mathbf{s}_{t,i} = (re_{t,i},\; p_{t,i},\; \iota_{t,i},\; u_{t,i},\; f_{t,i})
\end{equation}
where $re_{t,i}$ denotes relevance, $p_{t,i}$ denotes polarity, $\iota_{t,i}$ denotes intensity,
$u_{t,i}$ denotes uncertainty, and $f_{t,i}$ denotes forwardness.
All dimensions are elicited on a continuous scale via structured
prompts.

To aggregate article-level scores to the weekly level, we compute a relevance-weighted mean for each sentiment dimension:
\begin{equation}
    \bar{w}_{t} = \frac{\sum_{i \in \mathcal{A}_t} re_{t,i} \cdot 
    w_{t,i}}{\sum_{i \in \mathcal{A}_t} re_{t,i}},
    \quad w \in \{p, \iota, u, f\},
    \label{eq:aggregation}
\end{equation}
where $\mathcal{A}_t$ is the set of articles published in week $t$. Relevance scores serve as aggregation weights on the grounds that articles with stronger topical alignment to \wti{} markets should exert greater influence on the weekly signal. As relevance is already incorporated into the weighting scheme, no additional relevance filtering is applied. 

Following the weekly aggregated means, we additionally construct two dispersion features, the within-week standard deviation of polarity and the within-week standard deviation of uncertainty, to capture the degree of disagreement across articles within each week. A high standard deviation of polarity indicates that the market is simultaneously receiving conflicting signals, reflecting bullish opinion in some articles and bearish opinion in others. Similarly, a high standard deviation of uncertainty indicates that articles within the same week express markedly different levels of ambiguity, which may carry information about the informativeness of the prevailing news flow.

In addition to weekly level features, we construct first-difference momentum terms for polarity, uncertainty, and forwardness as follows: 
\begin{equation}
    \Delta \bar{w}_{t} = \bar{w}_{t} - \bar{w}_{t-1},
    \quad w \in \{p, u, f\},
    \label{eq:momentum}
\end{equation}

The rationale for including momentum is grounded in the behavioral finance literature, which documents that sentiment trends, rather than sentiment levels alone, carry predictive information about asset returns \citep{baker2006investor,uhl2015s}
A sustained shift in polarity, for instance, may signal an emerging consensus that is not yet reflected in prices, while a sudden reversal may indicate that the prevailing narrative is losing credibility.

We select these three features for momentum construction is due to their distinct temporal dynamics. Polarity captures the directional tone of market sentiment, whose trend is directly relevant to price momentum. Uncertainty momentum reflects whether market ambiguity is rising or falling week-on-week, a signal that is particularly informative around supply shocks and geopolitical events when the information environment is rapidly evolving. Forwardness momentum captures shifts in the degree to which news coverage is oriented toward future expectations versus past events, which we interpret as a proxy for changing market attention. 

In contrast, momentum terms for relevance and intensity are excluded. Relevance is related to news connection to the market and its week-on-week change does not carry an interpretable market meaning. Intensity, while useful as a weekly level feature, does not have a well-defined directional implication. An  increase in emotional strength across articles is not systematically associated with a particular price direction. 

In total, we construct 31 features as the candidate independent features and the full list is provided in  \ref{app:features}

\subsection{Prediction Model}
\label{SubsecModel}

We consider LightGBM \citep{ke2017lightgbm} as the prediction model, a gradient boosting framework well-suited for tabular dataset with moderate sample size and robust to feature scale differences \citep{grinsztajn2022tree,shwartz2022tabular}. 

Regarding feature selection, we consider six sets of different feature combinations as follows. 
\begin{itemize}
    \item \textit{Set AV Baseline}: features from AlphaVantage only, serving as the benchmark as it relies solely on the proprietary AlphaVantage score. 
    \item \textit{Set Tradition}:features from AlphaVantage and \finbert, representing a conventional multi-source combination without LLM-derived dimensions. 
    \item \textit{Set GPT}: features from \gpt only.
    \item \textit{Set Llama}: features from \llama only.
    \item \textit{Set LLM}: features from  \gpt and  \llama combined
    \item \textit{Set GPT-FinBert}: features from  \gpt and \finbert combined
\end{itemize}

We train separate LightGBM models for each feature set. To optimize  hyperparameters, we employ Optuna \citep{akiba2019optuna}, a hyperparameter optimization framework based on the Tree-structured Parzen Estimator (TPE) algorithm \citep{bergstra2011algorithms}. TPE constructs two probabilistic models $p(x)$ and $g(x)$ over the hyperparameter space, representing the densities of configurations that yield above- and below-threshold performance respectively, and selects candidates by maximizing the ratio $\frac{p(x)}{g(x)}$, which is equivalent to maximizing the Expected Improvement. This approach is more sample-efficient than grid search or random search, particularly in high-dimensional hyperparameter spaces. The whole tuning process is performed via time-series cross-validation with $K=5$ expanding windows, using Area Under the Receiver Operating Characteristic Curve (AUROC) as the optimization criteria. 

To evaluate model performance, we consider three metrics as follows: 
\begin{itemize}
    \item Area Under the Receiver Operating Characteristic Curve (AUROC): It summaries classification performance across all possible decision thresholds by plotting the true positive rate against the false positive rate. A value of 0.5 is equivalent to random guessing, while a value of 1.0 indicates perfect discriminative ability.
    \item Accuracy: It measures the proportion of weeks for which the predicted direction matches the realized direction. Unlike AUROC, it evaluates the binary classification decision at a fixed threshold rather than across all thresholds. Given that positive-return weeks is approximately $53.03\%$ of the sample, we set the classification threshold to match the empirical class distribution. 
    \item Information Coefficient (IC): \begin{equation}
\text{IC} = \rho_s\left(\hat{p}_t,, r_t\right) = 1 - \frac{6\sum d_t^2}{T(T^2-1)}
\label{eq:ic}
\end{equation}
where $\hat{p}_t$ is the predicted probability of a positive return in week $t$, $r_t$ is the realized week return, and $d_t$ is the difference in ranks between $\hat{p}_t$ and $r_t$. IC measures the rank correlation between predicted probabilities and actual returns, capturing whether weeks assigned higher predicted probabilities tend to realized higher returns. Unlike AUC and Accuracy, IC evaluates the quality of the full predicted distribution rather than the binary classification decision.
\end{itemize}

\subsection{SHAP-Based Interpretation of Feature Importance}
To interpret the contribution of sentiment features to the predictive model,
we use SHapley Additive exPlanations (SHAP) \citep{lundberg2017unified} for the fitted LightGBM classifier. SHAP is a post-hoc explainability framework grounded in cooperative game theory. It attributes a model’s prediction to individual input features through Shapley values, which quantify each feature's margnial contribution to the model output. 

Formally, the Shapley value for feature $j$ is defined as

\begin{equation}
\phi_j =
\sum_{S \subseteq F \setminus \{j\}}
\frac{|S|!(|F|-|S|-1)!}{|F|!}
\left[
f_{S \cup \{j\}}(x_{S \cup \{j\}})
-
f_S(x_S)
\right]
\end{equation}

where $F$ denotes the full set of features, and $S$ represents a subset of
features excluding feature $i$. The term $f_S(x_S)$ denotes the model
output when only the feature subset $S$ is observed, while
$f_{S \cup \{j\}}(x_{S \cup \{j\}})$ represents the prediction when feature
$j$ is added to the subset. The difference between the two terms therefore captures the marginal contribution of feature $j$, averaged over all possibile feature coalitions.

SHAP provides both local and global interpretability. At the local
level, the value $\phi_{i,j}$ quantifies the contribution of
feature $j$ to the prediction for observation $i$. A positive SHAP value
indicates that the feature pushes the prediction toward the positive class, while a negative value indicates a downward contribution. At the global level, the overall feature importance is evaluated by the mean absolute SHAP value across all observations:
\begin{equation}
\Phi_j =
\frac{1}{N}
\sum_{i=1}^{N}
|\phi_{i,j}|
\end{equation}
where $N$ is the total number of observations. Features with larger mean absolute SHAP values exert greater influence on the model's predictions and are therefore interpreted as more important drivers of weekly \wti{} futures return direction.

\section{Results}
\label{SecResults}

\subsection{Inter-Model Agreement}
\label{SocreAnalysis}
\begin{table}[htbp]
\centering
\caption{Distributional Properties of Polarity Scores Across Models}
\label{tab:polarity_stats}
\begin{tabular}{lcccc}
\hline
 & \textbf{GPT-4o} & \textbf{Llama 3.2} & \textbf{FinBERT} & \textbf{AlphaVantage} \\
\hline
Mean   & 0.2952 & 0.1109 & 0.3050 & 0.2177 \\
Stand Deviation    & 0.4323 & 0.3458 & 0.6801 & 0.2818 \\
Min    & -0.9000 & -1.0000 & -0.9772 & -0.9144 \\
$Q25\%$     & 0.0000 & 0.0000 & 0.0000 & 0.0368 \\
Median & 0.3000 & 0.0000 & 0.6491 & 0.2758 \\
$Q75\%$     & 0.7000 & 0.2000 & 0.9000 & 0.4218 \\
Max    & 0.9000 & 0.9000 & 0.9616 & 0.9419 \\
\hline
\end{tabular}
\end{table}

Table~\ref{tab:polarity_stats} summarizes the distribution of polarity scores across approaches. All approaches yield positive mean polarity scores, ranging from 0.11 for \llama to 0.31 \finbert, suggesting that energy news carries a modest positive tone on average. Among those approaches, \llama exhibits the most conservative scoring pattern, with both the $25\%$ quartile and the median as zero. In fact, 3608 out of 8639 news ($41.7\%$) receives a polarity score of zero under \llama.

\finbert exhibits a substantially different distribution from the LLM-based measures. Its median polarity is 0.65 and the third quartile is 0.90, indicating that many articles are assigned strongly positive scores, although the distribution also includes a long negative tail, with a minimum of -0.98. This pattern is consistent with the distinct architecture of this model. As a classification model, \finbert produces polarity scores derived from class probabilities, which tend to place greater mass near the extremes. By contrast, \gpt and \llama produce smoother and more graduated polarity assessments.

\begin{figure}[htbp]
    \centering
    \includegraphics[width=0.6\textwidth]{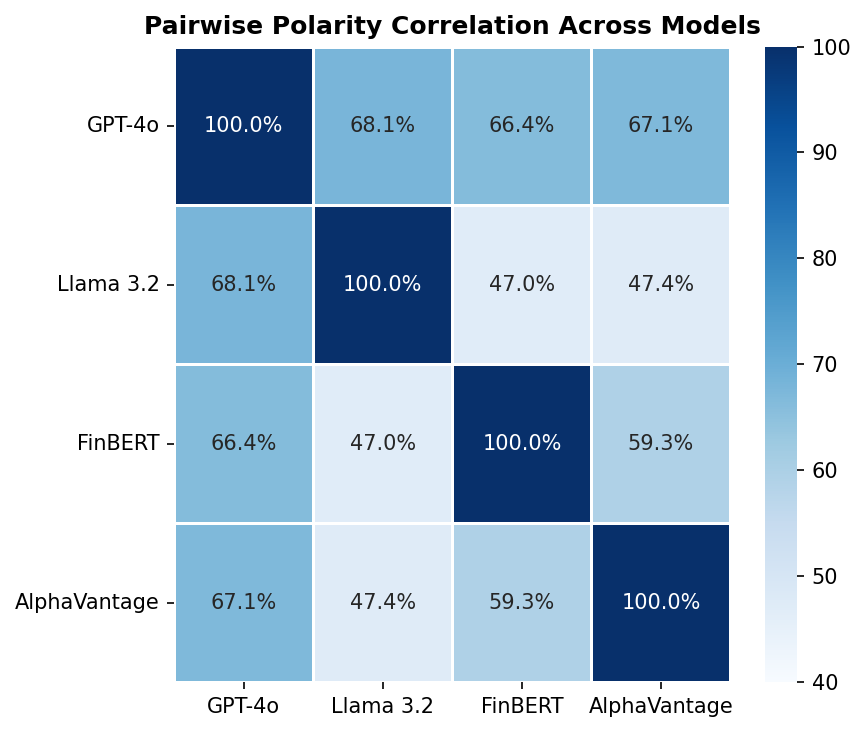}
    \caption{Pairwise Polarity Correlation Across Models}
    \label{fig:polarity_heatmap}
\end{figure}

Figure~\ref{fig:polarity_heatmap} presents the pairwise Pearson correlation among polarity scores generated by the four sentiment sources. All correlations are positive and moderate, ranging from $47\%$ to $68\%$. This pattern indicates that the models share a common directional signal, but also retain substantial cross-model variation.\gpt and \llama exhibit the highest correlation ($68.1\%$), which is consistent with the fact that both are instruction-tuned language models evaluated under the same prompt design. By contrast, \llama shows the weakest agreement with \finbert ($47.0\%$) and AlphaVantage ($47.4\%$), suggesting that the smaller language model captures a somewhat different component of sentiment from both the larger LLM and the more conventional approach. Overall, the moderate correlation motivates for treating these sentiment signals as complementary inputs rather than redundant proxies, leading the multi-model feature construction in Section~\ref{SubsecModel}.

\begin{figure}[htbp]
    \centering
    \includegraphics[width=\textwidth]{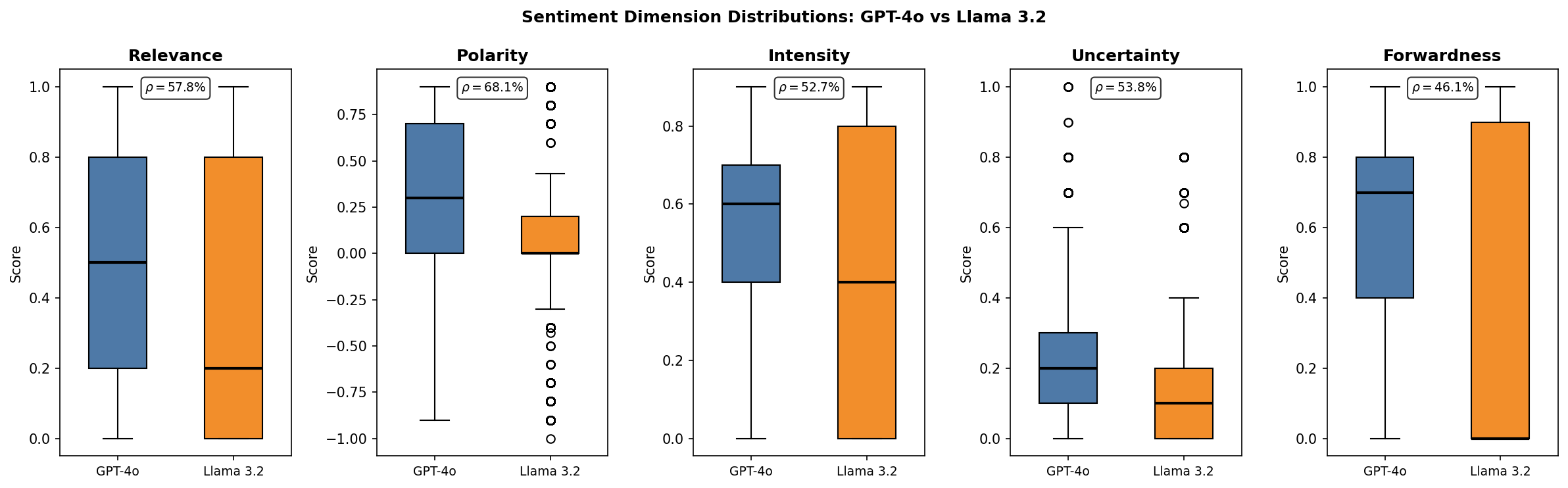}
    \caption{Sentiment Dimension Distributions: GPT-4o vs Llama 3.2}
    \label{fig:dimension_boxplot}
\end{figure}

Figure~\ref{fig:dimension_boxplot} compares the weekly aggregated sentiment dimensions by \gpt and \llama. \gpt consistently yields higher average scores  across all five dimensions, suggesting greater sensitivity to sentiment-related cues in energy news. Polarity displays the strongest alignment with correlation $52.7\%$, indicating the two models capture broadly similar directional sentiment. Conversely, the differences become more pronounced for other dimensions, particularly forwardness and intensity, where \llama exhibits substantially greater dispersion. Taken together, these findings indicate that \gpt and \llama are not interchangeable sentiment sources. Instead, each appears to preserve distinct information, provideing empirical support for combining them in the predictive feature set.

\subsection{Directional Prediction Performance}

\begin{figure}[htbp]
    \centering
    \includegraphics[width=\textwidth]{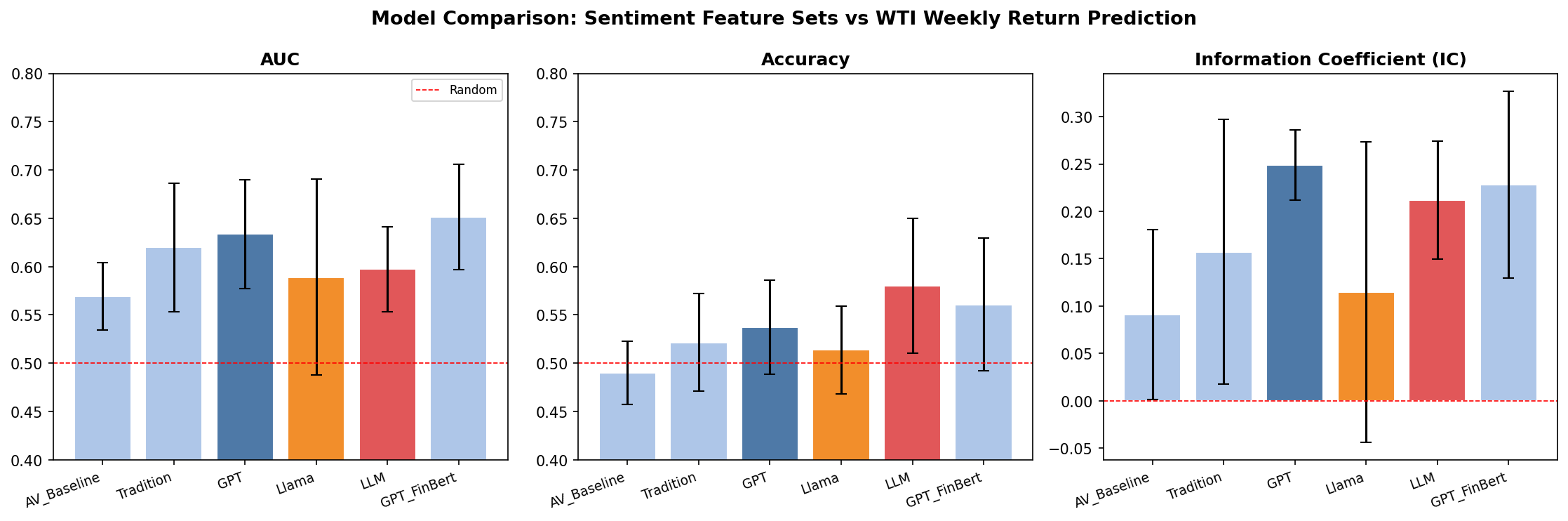}
    \caption{Model Comparisons}
    \label{fig:model_compare}
\end{figure}

Figure~\ref{fig:model_compare} and Table~\ref{tab:model_results} report the out-of-sample performance of the six feature sets  under five-fold time series cross-validation. All models achieve AUC values above the benchmark baseline of 0.5, indicating that sentiment-based features contain meaningful predictive information for \wti{} weekly return direction.
The AV Baseline, sourced from AlphaVantage proprietary sentiment scores, has an AUC of 0.569 and an IC of 0.091, making it weakest performance. Expanding the baseline with \finbert signals improves the performance with AUC of 0.620 and IC of 0.157. This improvement suggests that combining multiple sentiment sources adds incremental predictive information.

Among the LLM-based sources, \gpt alone achieves a mean AUC of 0.634 and the highest mean IC of 0.249, outperforming both \llama and the LLM Ensemble. This pattern is consistent with the earlier score distribution evidence in Section \ref{SocreAnalysis}, where \llama displays less stable sentiment scoring across all the five dimensions. Notably, combining \gpt and \llama does not improve upon \gpt alone, suggesting that signals from \llama may introduce noise rather than complementary information.

The highest AUC is obtained by the combination of \gpt and \finbert. Although its IC is slighly lower than that of \gpt alone (0.228 versus 0.249), the difference is small and should be interpreted cautiously. Overall, these results suggest that \finbert signal provide additional information helpful for classification performance, which is not captured by \gpt alone.

\begin{table}[htbp]
\centering
\caption{Cross-Validation Prediction Performance Across Feature Sets}
\label{tab:model_results}
\begin{tabular}{lccccccc}
\hline
\textbf{Model} & \multicolumn{2}{c}{\textbf{AUC}} & \multicolumn{2}{c}{\textbf{Accuracy}} & \multicolumn{2}{c}{\textbf{IC}} \\
\cmidrule(lr){2-3} \cmidrule(lr){4-5} \cmidrule(lr){6-7}
 & Mean & Std & Mean & Std & Mean & Std \\
\hline
AV Baseline   & 0.5694 & 0.0349 & 0.4902 & 0.0328 & 0.0910 & 0.0894 \\
Tradition   & 0.6200 & 0.0666 & 0.5216 & 0.0505 & 0.1571 & 0.1398 \\
GPT-4o        & 0.6336 & 0.0563 & 0.5373 & 0.0490 & 0.2490 & 0.0372 \\
Llama 3.2     & 0.5892 & 0.1013 & 0.5137 & 0.0454 & 0.1147 & 0.1586 \\
LLM Ensemble  & 0.5974 & 0.0439 & 0.5804 & 0.0697 & 0.2118 & 0.0620 \\
GPT + FinBERT & 0.6515 & 0.0546 & 0.5608 & 0.0686 & 0.2283 & 0.0987 \\
\hline
\end{tabular}
\end{table}

\subsection{SHAP Feature Importance}

\begin{figure}[htbp]
    \centering
    \includegraphics[width=\textwidth]{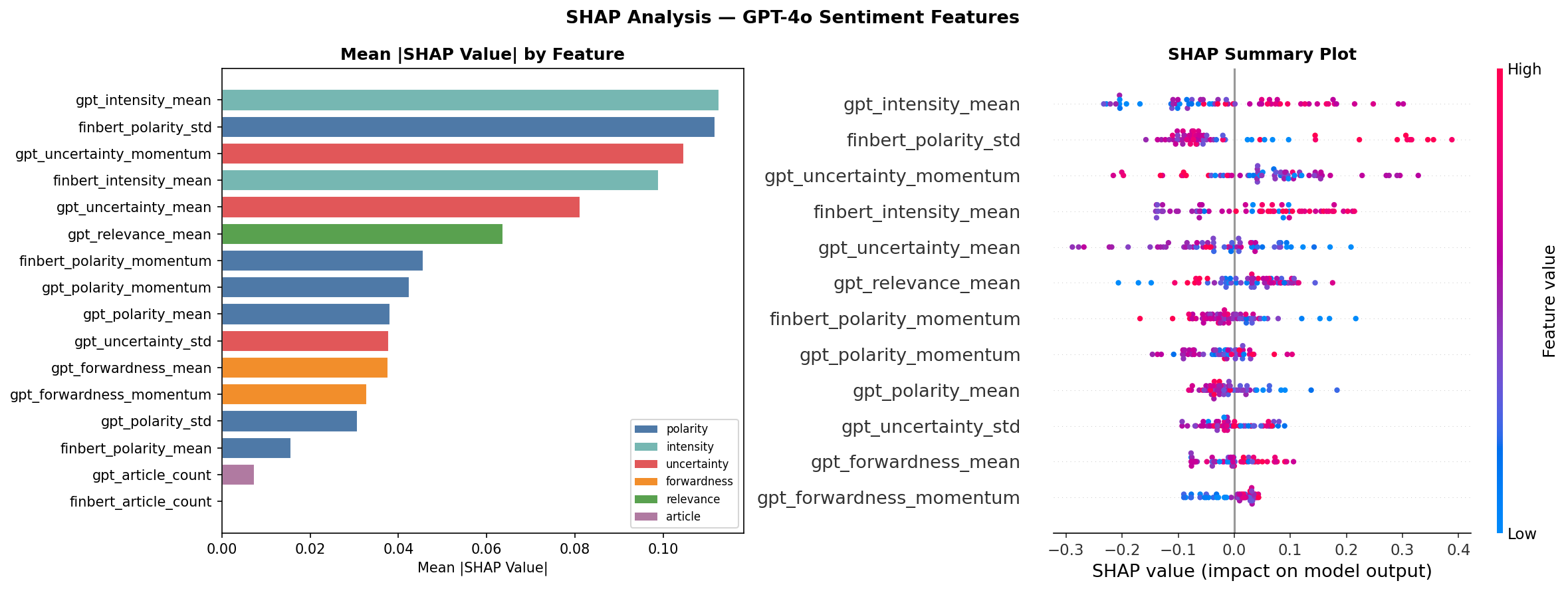}
    \caption{SHAP Analyais}
    \label{fig:shap}
\end{figure}

Figure~\ref{fig:shap} presents the SHAP-based feature importance analysis for the GPT + FinBERT model, computed on the held-out test set. The left panel reports mean absolute SHAP values, while the right panel shows the direction and magnitude of each feature's contribution to the model output.The two most important features are \texttt{gpt\_intensity\_mean} and \texttt{finbert\_polarity\_std}, followed by \texttt{gpt\_uncertainty\_momentum}. The prominence of intensity and uncertainty, rather than polarity, suggests that dimensions beyond directional sentiment contain meaningful predictive information for \wti{} weekly return direction. Notably, \texttt{finbert\_polarity\_std} ranks second overall, suggesting that cross-article sentiment variation is the primary channel through which \finbert contributes to the ensemble. This findings is consistent with the multi-model design, as \gpt and \finbert appear to capture distinct aspects of the information environment. However, \texttt{gpt\_article\_count} and \texttt{finbert\_article\_count} have negligible contribution, indicating that article volume alone adds little incremental predictive content once sentiment features are included.

The SHAP summary plot provides additional insight into the direction of feature effects. High values of \texttt{gpt\_intensity\_mean} are associated with positive SHAP contributions, indicating that weeks with stronger overall intensity are more likely to be classified as up-weeks. In contrast, higher \texttt{gpt\_uncertainty\_mean} is associated with negative SHAP values, suggesting that elevated aggregate uncertainty pushes predictions to down-weeks. Interestingly, \texttt{gpt\_uncertainty\_momentum} exhibits the opposite pattern, with rising uncertainty associated with positive SHAP contributions. This observation suggests that the level and the change in uncertanity may convey different predictive information to the model.


\section{Conclusions}
\label{SecConclusion}

This paper examines whether multi-dimensional sentiment signals extracted by large language models provide incremental predictive content for weekly \wti{} crude oil futures returns. Using energy-sector news articles from 2020 to 2025, we construct five sentiment dimensions, including relevance, polarity, intensity, uncertainty, and forwardness, based on \gpt and \llama, and benchmark them against \finbert and the AlphaVantage sentiment measures.

Our findings are twofold. First, the combination of \gpt and \finbert delivers the strongest overall predictive performance among the feature sets considered, indicating that richer model-based sentiment representations contain incremental forecasting information. Second, the SHAP analysis shows that intensity and uncertainty-related features rank among the most important predictors, suggesting that the predictive content of news sentiment extends beyond polarity alone.

From a practical perspective, the results suggest that commodity investors, trading desks, and risk managers may benefit from supplementing traditional sentiment monitoring with LLM-based measures of uncertainty and sentiment intensity. More broadly, the evidence indicates that multi-dimensional sentiment extraction can improve the informational value of news-based signals in commodity return prediction.

This study has several limitations. The analysis focuses on a single commodity and a single news source, and the weekly forecasting horizon may not generalize to higher-frequency settings. Future research could extend the framework to other energy commodities, incorporate additional data sources such as shipping or satellite information, and evaluate performance at shorter horizons. The choice of  \gpt and \llama are based to a cost-performance tradeoff that merits further investigation as LLM models continue to improve.

\section*{Acknowledgments}
During the preparation of this manuscript, the authors used ChatGPT for the purposes of manuscript polishing only. The authors have reviewed and edited the manuscript and take full responsibility for the content of this publication.

\appendix

\section{System Prompts}
\label{app:prompts}

Below is the system prompt designed for five dimension extraction using LLM models. 

\begin{lstlisting}[breaklines=true, basicstyle=\small\ttfamily]
You are a financial news analyst specializing in energy markets.

Analyze the following news article and return a JSON object with exactly these fields:

{
  "relevance": float,      // Relevance to energy markets (oil, gas, coal, energy policy): 0.0 (unrelated) to 1.0 (directly relevant)
  "polarity": float,       // Overall sentiment toward energy markets: -1.0 (very negative) to +1.0 (very positive)
  "intensity": float,      // Strength of sentiment: 0.0 (neutral/weak) to 1.0 (very strong)
  "uncertainty": float,    // Degree of uncertainty/ambiguity: 0.0 (certain) to 1.0 (very uncertain)
  "forwardness": float     // Forward-looking vs backward-looking: 0.0 (past events) to 1.0 (future outlook)
}

Rules:
- Evaluate relevance FIRST. If relevance < 0.1, set all other fields to null.
- polarity and intensity are independent: a strongly worded negative article has polarity=-0.9, intensity=0.9
- uncertainty reflects hedge words: "may", "could", "uncertain", "at risk", "unclear"
- forwardness reflects future tense, forecasts, expectations, projections vs reported past events
- Return only the JSON object, no explanation.
\end{lstlisting}

\section{Feature List}
\label{app:features}
\setlength{\LTleft}{-2cm}
\begin{longtable}{llcp{5.5cm}}
\caption{Full Feature List with Descriptions and Ranges}
\label{tab:features} \\
\hline
\textbf{Feature Name} & \textbf{Model} & \textbf{Range} & \textbf{Description} \\
\hline
\endfirsthead
\hline
\textbf{Feature Name} & \textbf{Model} & \textbf{Range} & \textbf{Description} \\
\hline
\endhead
\hline
\endfoot

\texttt{gpt\_article\_count}        & GPT-4o  & {(}4, 394{)} & Number of sampled articles in the week \\
\texttt{gpt\_relevance\_mean}       & GPT-4o  & {(}0.18, 0.8{)}     & Mean relevance to energy markets, used as aggregation weight \\
\texttt{gpt\_polarity\_mean}        & GPT-4o  & {(}-0.35, 0.71{)}    & Relevance-weighted mean sentiment polarity \\
\texttt{gpt\_intensity\_mean}       & GPT-4o  & {(}0.42, 0.72{)}     & Relevance-weighted mean sentiment intensity \\
\texttt{gpt\_uncertainty\_mean}     & GPT-4o  & {(}0.13, 0.41{)}     & Relevance-weighted mean degree of uncertainty \\
\texttt{gpt\_forwardness\_mean}     & GPT-4o  & {(}0.33, 0.83{)}     & Relevance-weighted mean forward-looking orientation \\
\texttt{gpt\_polarity\_std}         & GPT-4o  & {(}0.22, 0.65{)}     & Cross-article dispersion of polarity within the week \\
\texttt{gpt\_uncertainty\_std}      & GPT-4o  & {(}0.05, 0.26{)}     & Cross-article dispersion of uncertainty within the week \\
\texttt{gpt\_polarity\_momentum}    & GPT-4o  & {(}-0.82, 0.73{)}    & Week-on-week change in polarity mean \\
\texttt{gpt\_uncertainty\_momentum} & GPT-4o  & {(}-0.16, 0.17{)}    & Week-on-week change in uncertainty mean \\
\texttt{gpt\_forwardness\_momentum} & GPT-4o  & {(}-0.31, 0.39{)}    & Week-on-week change in forwardness mean \\
\hline

\texttt{llama\_article\_count}        & Llama 3.2 & ${(}4, 394{)}$ & Number of sampled articles in the week \\
\texttt{llama\_relevance\_mean}       & Llama 3.2 & {(}0, 0.67{)}     & Mean relevance to energy markets \\
\texttt{llama\_polarity\_mean}        & Llama 3.2 & {(}-0.4, 0.7{)}    & Relevance-weighted mean sentiment polarity \\
\texttt{llama\_intensity\_mean}       & Llama 3.2 & {(}0.36, 0.9{)}     & Relevance-weighted mean sentiment intensity \\
\texttt{llama\_uncertainty\_mean}     & Llama 3.2 & {(}0.03, 0.4{)}     & Relevance-weighted mean degree of uncertainty \\
\texttt{llama\_forwardness\_mean}     & Llama 3.2 & {(}0.02, 1{)}     & Relevance-weighted mean forward-looking orientation \\
\texttt{llama\_polarity\_std}         & Llama 3.2 & {(}0, 0.55{)}     & Cross-article dispersion of polarity within the week \\
\texttt{llama\_uncertainty\_std}      & Llama 3.2 & {(}0, 0.29{)}     & Cross-article dispersion of uncertainty within the week \\
\texttt{llama\_polarity\_momentum}    & Llama 3.2 & {(}-0.65, 0.82{)}    & Week-on-week change in  polarity mean \\
\texttt{llama\_uncertainty\_momentum} & Llama 3.2 & {(}-0.27, 0.25{)}    & Week-on-week change in  uncertainty mean \\
\texttt{llama\_forwardness\_momentum} & Llama 3.2 & {(}-0.57, 0.76{)}    & Week-on-week change in forwardness mean \\
\hline

\texttt{finbert\_article\_count}        & FinBERT & {(}4, 394{)} & Number of sampled articles in the week \\
\texttt{finbert\_polarity\_mean}        & FinBERT & {(}-0.44, 0.82{)}    & Equal-weighted mean sentiment polarity \\
\texttt{finbert\_polarity\_std}         & FinBERT & {(}0.12, 0.87{)}     & Cross-article dispersion of polarity within the week \\
\texttt{finbert\_intensity\_mean}       & FinBERT & {(}0.17, 0.88{)}     & Equal-weighted mean sentiment intensity \\
\texttt{finbert\_polarity\_momentum}    & FinBERT & {(}-0.77, 0.86{)}    & Week-on-week change in polarity mean  \\
\hline

\texttt{av\_article\_count}        & AlphaVantage & {(}4, 394{)} & Number of articles in the week \\
\texttt{av\_polarity\_mean}        & AlphaVantage & {(}-0.17, 0.41{)}    & Equal-weighted mean AV sentiment score \\
\texttt{av\_polarity\_std}         & AlphaVantage & {(}0.12, 0.47{)}     & Cross-article dispersion of AV sentiment score \\
\texttt{av\_polarity\_momentum}    & AlphaVantage & {(}-37, 0.49{)}    & Week-on-week change in polarity mean \\

\end{longtable}










\bibliographystyle{plainnat}
\bibliography{sentiment}

\end{document}